\documentstyle[prb,epsfig,aps,eqsecnum]{revtex}

\begin{document}
\title{Probing Ion-Ion and Electron-Ion Correlations in Liquid Metals
within the Quantum Hypernetted Chain Approximation} 
\author{J.A. Anta\protect{\cite{Juan}}}
\address{ Physical and Theoretical Chemistry Laboratory, Oxford
University, South Parks Road, Oxford OX1 3QZ, UK} \author{A.A. Louis}
\address{ Department of Chemistry, Cambridge University,
Lensfield Rd, Cambridge CB2 1EW, UK} \date{\today} \maketitle
\begin{abstract}
We use the Quantum Hypernetted Chain Approximation (QHNC) to calculate
the ion-ion and electron-ion correlations for liquid metallic Li, Be,
Na, Mg, Al, K, Ca, and Ga. We discuss trends in electron-ion structure
factors and radial distribution functions, and also calculate the
free-atom and metallic-atom form-factors, focusing on how bonding
effects affect the interpretation of X-ray scattering experiments,
especially experimental measurements of the ion-ion structure factor
in the liquid metallic phase.

\vspace*{4pt}
\end{abstract}
\vspace{20pt}


\section{Introduction}

Liquid metals are complex binary fluids consisting of ions in a sea of
conduction electrons. While the ions can usually be treated classically, 
the electrons are typically degenerate and must be treated
quantum-mechanically. Liquids are differentiated from gases by
non-trivial structure at the level of two-body correlation functions;
they are generally close in density to solid phases.  For
two-component systems these correlation functions are defined in
$k$-space as:
\begin{equation}\label{eq1.1} 
S_{\alpha \beta}(k) = \frac{<\hat{\rho}_\alpha({\bf k})
\hat{\rho_\beta}(-{\bf k})>}{(N_\alpha N_\beta)^{1/2}} - (N_\alpha
N_\beta)^{1/2} \delta_{k,0}.
\end{equation}
The $S_{\alpha \beta}(k)$ are referred to as static structure factors
and the operator
\begin{equation}\label{eq1.2}
\hat{\rho}_\alpha({\bf k}) = \sum_{i=1}^{N_\alpha} e^{i{\bf k}\cdot
{\bf r}_{i \alpha }},
\end{equation} is the Fourier transform of the one-particle density operator of
component $\alpha$. The indices $\alpha$ and $\beta$ refer to ions,
($I$), or valence electrons, ($e$).
The structure factors $S_{\alpha \beta}(k)$ can be related to the
so-called radial distribution functions $g_{\alpha \beta}(r)$ by:
\begin{equation}\label{eq1.4}
S_{\alpha \beta}(k) = \delta_{\alpha \beta} + \left( \rho_\alpha
\rho_\beta\right)^{\frac{1}{2}} \int_V d {\bf r} e^{i {\bf k}\cdot
{\bf r}} \left[ g_{\alpha \beta} (r) -1 \right],
\end{equation}
where the $\rho_i$ are the homogeneous average densities.

The determination of the ion-ion structure factor $S_{II}(k)$ and
the electron-electron structure factor $S_{ee}(k)$ are interesting
problems in their own right (one largely quantum mechanical, the other
largely classical), and have been the focus of much research: the
$S_{II}(k)$ because of their experimental accessibility; the
$S_{ee}(k)$ (with the ions usually smeared into a rigid neutralising
background) because of the importance of the electron
fluid\cite{foot1}.  In contrast, the electron-ion structure factor
$S_{eI}(k)$ has received considerably less attention, partially
because it is hard to measure, partially because its exact physical
relevance remains largely unexplored and unknown, and partially
because it includes both the physics of the ions {\em and} the physics
of the electrons, each of which is traditionally treated with its own
set of theoretical techniques.

One of the simplest ways to treat the valence electrons in a liquid
metal is in a linear response formalism using a local
pseudo-potential\cite{Ashc78}.  In fact, linear response has been
shown to be much more accurate than one would na\"{\i}vely expect, a result
which stems in part from a recently discovered interference effect
between an atomic lengthscale, the inverse ionic length, and an
electronic lengthscale, twice the Fermi wave vector $2
k_F$\cite{Loui98b}.  This interference effect significantly reduces
the magnitude of the non-linear response terms at the normal densities
of most liquid metals.  The electron-ion correlations emerge when the
induced linear response electron density is combined with standard
liquid state techniques to treat the
ions\cite{Cusa76,Boul97,Wax97,Marc98}.  This approach is easy to
implement, can in some cases be remarkably accurate, and can explain
the qualitative trends in the shape of the electron-ion structure
factor $S_{eI}(k)$ for metallic liquids across the periodic
table\cite{Loui98b}.  The main obstacles to higher accuracy lie in the
uncertainty over the exact (local) pseudo-potential, especially when
non-local effects are important\cite{Lai98}, and also in the neglect of
non-linear electron response and of ion-ion correlation effects on the
induced electron densities\cite{Loui98a,Anta98}.

The development of {\em ab initio} simulation techniques based on
Density Functional Theory (DFT) for the electrons\cite{Kohn65}, and
molecular-dynamics on the adiabatic electronic potential energy
surface for the ions\cite{Car85}, provide probably the most accurate
and well-tested approach to electron-ion structure.  However, the
drawback of these methods is their computational cost; in practice
only relatively small system sizes can be investigated and so far only
results for Mg and Bi electron-ion correlations have been
published\cite{deWi95}.  The related Orbital-Free {\em ab initio}
molecular dynamics method (OF-AIMD)\cite{Pear93} allows larger system
sizes and significantly longer simulation times, and has been
successfully applied to the electron-ion correlations of
Li, Na, Mg, and Al\cite{Anta98,Anta99}, but the computational
cost is still rather large.

An alternative approach  is the Quantum Hypernetted Chain
(QHNC) method of Chihara\cite{Chih78}, which self-consistently
combines integral equation techniques from the theory of simple
liquids with a Kohn-Sham type treatment for the electrons.  The QHNC
treats the electrons and ions on essentially equal footing, does not
require a pseudo-potential approximation, and is computationally
relatively cheap.  Ion-ion and electron-ion correlations emerge in the
thermodynamic limit -- there are no finite size effects.

In section II, we derive the basic form of the QHNC approximation by
focusing first on the exact Quantum Ornstein Zernike (QOZ) equations
in section II.A, and then outlining the approximations needed to
derive the QHNC approximation in section II.B.  The numerical
implementation of the QHNC is detailed in the appendix.

In section III, we describe the ion-ion and electron-ion correlations
that emerge from the QHNC for our set of metals: Li, Be, Na, Mg, Al,
K, Ca, and Ga.

Even though the valence electron distributions are changed in a bonded
environment, X-ray scattering off liquid metals has traditionally been
interpreted with a free-atom form factor.  In section IV, we describe
the difference between extracting ion-ion structure in X-ray
scattering with a free-atom form factor and extracting ion-ion
structure with a metallic-atom form factor.  The effects of bonding on
the coherent X-ray scattering intensity may be measured by comparing
X-ray and neutron scattering determinations of the ion-ion structure
factor $S_{II}(k)$.  However, experiments and theory have yet to
converge on this issue.

Finally, we present some concluding remarks in section V, and describe
some details related to the numerical implementation of the QHNC in 
the Appendix.

\section{Quantum Hypernetted Chain Approximation (QHNC)}

\subsection{Quantum Ornstein Zernike Relations}

The Quantum Ornstein Zernike Relations (QOZ) for a two-component
system are most naturally derived in the context of density functional
theory (DFT)\cite{Loui98a,Xu98}.  First we define the Helmholtz
free-energy for a two-component system, which is a unique functional
of the two one-body density profiles\cite{Merm65}:
\begin{equation}\label{eq2.1}
F[\rho_1,\rho_2] = F_1^{id}[\rho_1] + F_2^{id}[\rho_2] + 
F^{ex}[\rho_1,\rho_2].
\end{equation}
The functional is split in the usual way between ideal
(non-interacting) and excess (interacting) parts. We then introduce the
external potential field:
\begin{equation}\label{eq2.2}
\Psi_\alpha({\bf r}) = \mu_\alpha - \phi_\alpha({\bf r}),
\end{equation}
which is defined in terms of the chemical potential $\mu_\alpha$ of
species $\alpha$ and the external potential, $\phi_\alpha({\bf r})$
which acts on species $\alpha$ only.  A Legendre transform with
respect to these external fields obtains the grand potential:
\begin{equation}\label{eq2.3}
\Omega[\Psi_1,\Psi_2] = F[\rho_1, \rho_2] + 
\int d{\bf r}\rho_1({\bf r}) \Psi_1({\bf r}) + 
 \int d{\bf r}\rho_2({\bf r}) \Psi_2({\bf r}),
\end{equation}
 which is in turn a unique functional of the two external potential
fields $\Psi_1$ and $\Psi_2$.  

The first two functional derivatives of the Helmholtz free-energy functional
w.r.t. the one-particle densities are:
\begin{equation}\label{eq2.5a}
\frac{\delta F}{\delta\rho_\alpha({\bf r})}  = 
 	\Psi_\alpha({\bf r}),
\end{equation}
and
\begin{equation}\label{eq2.5b}
\frac{\delta^2 F}
{\delta\rho_\alpha({\bf r})\delta\rho_\beta({\bf r}^\prime)}  = 
\frac{\delta\Psi_\alpha({\bf r})}{\delta\rho_\beta({\bf r}^\prime)} 
= \chi^{-1}_{\alpha\beta}({\bf r},{\bf r}^\prime)
\end{equation}

The first two derivatives of the grand
potential functional w.r.t. the external potential field are:
\begin{equation}\label{eq2.4a}
\frac{\delta\Omega}{\delta\Psi_\alpha({\bf r})} 
= \rho_\alpha({\bf r}),
\end{equation}
and
\begin{equation}\label{eq2.4b} 
\frac{\delta^2\Omega}
{\delta\Psi_\alpha({\bf r})\delta\Psi_\beta({\bf r}^\prime)} =
\frac{\delta\rho_\alpha({\bf r})}{\delta\Psi_\beta({\bf r}^\prime)} =
\chi_{\alpha\beta}({\bf r},{\bf r}^\prime),
\end{equation}
which defines the {\em susceptibility matrix} or matrix of the linear
response functions $\chi_{\alpha\beta}({\bf r},{\bf r}^\prime)$.
Thus the two 2nd functional derivatives are each others' functional
inverse, a natural consequence of having two  generating functionals
linked by a Legendre transform\cite{Zinn89}.

The direct correlation functions $C_{\alpha \beta}({\bf r,r'})$ of
an arbitrary two-component mixture are defined in the usual way as
functional derivatives of the excess free energy\cite{foot2}:
\begin{equation}\label{eq2.6}
\frac{-1}{\beta} C_{\alpha \beta}({\bf r},{\bf r'}) = \frac{\delta^2
F^{ex}}{\delta \rho_\alpha({\bf r}) \delta \rho_\beta({\bf r'})}.
\end{equation}
If we then define $(\chi^{(0)}_{\alpha \beta})^{-1}$ as the inverse
susceptibility matrix of the ideal system, we arrive, by combining
equations (\ref{eq2.1}), (\ref{eq2.5b}) and (\ref{eq2.6}), at the
following relationship between two 2-by-2 matrices:
\begin{equation}\label{eq2.7}
\frac{-1}{\beta} C_{\alpha \beta}  = (\chi_{\alpha \beta})^{-1} - 
(\chi^0_{\alpha \beta})^{-1},
\end{equation}
the {\bf Quantum Ornstein Zernike Relations (QOZ)}.  They follow from
simple properties of the two free-energy functionals and in this form
they are valid for {\em any} two-component inhomogeneous quantum
system (the generalisation to more than two components is
straightforward).  In the homogeneous limit
the direct correlation functions of Eq.~(\ref{eq2.6}) reduce to the
usual direct correlation functions first introduced by Ornstein and
Zernike\cite{Orns14,Hans86}, and it is in this sense that we will be using
them throughout the rest of this paper.

For classical species, the fluctuation-dissipation theorem relates
the response functions to density-density correlation functions\cite{Kubo64}:
\begin{equation}\label{eq2.8a}
\lim_{\hbar \rightarrow 0}\chi_{\alpha \beta}(k,0) 
= -\beta(\rho_\alpha \rho_\beta)^{1/2} S_{\alpha \beta}(k),
\end{equation} written here for a homogeneous system and in terms
of the structure factors defined in Eq.~(\ref{eq1.1}).  For a liquid
 metal, where the ions are viewed as classical but the electrons 
 quantum-mechanical, inverting the matrix in the QOZ relations of
 Eq.~(\ref{eq2.7}), and applying the fluctuation-dissipation theorem of
 Eq.~(\ref{eq2.8a}) for $\chi_{II}(k)$ and $\chi_{eI}(k)$ results in:
\begin{eqnarray}\label{eq2.8}
S_{II}(k) 
& = & 
[1+\chi^{(0)}_{ee}(k)C_{ee}(k)/\beta]/D(k) \nonumber \\
S_{eI}(k)
& = & 
-\sqrt{\frac{\rho_I}{\rho_e}}\chi^{(0)}_{ee}(k)(C_{eI}(k)/\beta)/D(k) 
\nonumber \\
\chi_{ee}(k) & = & \chi^{(0)}_{ee}(k)
  \left(1-\rho_I C_{II}(k)\right)/D(k) 
	\nonumber \\
D(k) & = & 
     \left[1-\rho_IC_{II}(k)\right]
     \left[1+\chi^{(0)}_{ee}(k)C_{ee}(k)/\beta\right]
	+\rho_I\chi^{(0)}_{ee}(k)|C_{eI}(k)|^2/\beta,
\end{eqnarray}
where $\chi^{(0)}_{ee}(k)$ is the well known {\em Lindhard
function}\cite{Ashc76}, the response function of the non-interacting
electron gas.  In the limit that {\em both} species are classical, the
QOZ relations reduce to the usual classical two-component
Ornstein-Zernike relations\cite{Hans86}.

The QOZ relations for a liquid metal appear to have been first derived
by Chihara\cite{Chih76}.  Later Ichimaru {\em et.\ al.}\cite{Ichi85}
derived similar equations from a two-component linear response
formulation.  The two formulations are equivalent if the definitions
of the direct correlation functions of Eq.~(\ref{eq2.6}) are linked in the
usual way to the local field factors $G_{\alpha \beta}(k)$:
\begin{equation}\label{eq2.10}
\frac{C_{\alpha \beta}(k)}{\beta} = -V_{\alpha \beta}(k)(1 -
 G_{\alpha \beta}(k)),
\end{equation}
where $V_{\alpha \beta}(k)$ is the direct potential between species.

\subsection{Quantum Hypernetted Chain Approximation}

To solve the QOZ relations for a liquid metal we recast them into a
slightly different form using two steps\cite{Loui98a}: The first step
is to use the Percus trick\cite{Perc62} to relate the
{\em homogeneous} two-body pair-correlation functions to the
one-body {\em inhomogeneous} density around one particle fixed at the
origin.  For the electron-ion pair-correlation function we fix an ion
at the origin to find:
\begin{equation}\label{eq2.11}
g_{eI}({\bf 0,r}) = \frac{\rho_e({\bf r}|I)}{\rho_e},
\end{equation}
where $\rho_e({\bf r}|I)$ is the (interacting) valence electron
density induced by one ion at the origin.  A similar relationship
holds for the ion-ion pair-correlation function, but for the
electron-electron pair correlation function the Percus trick cannot be
used in this form; one cannot ``fix'' an electron at the origin.

The second step follows  the basic ideas of the Kohn-Sham
scheme\cite{Kohn65}, namely that there exists a local single-particle
external potential $v^{\text{eff}}({\bf r})$ which will induce in a {\em
non-interacting} system the same one-particle density $\rho({\bf r})$
as is found in the full {\em interacting} system.  This idea holds
both for quantum as well as for classical systems.  The external
effective potential felt by species $\alpha$ when species $\beta$ is
fixed at the origin follows directly from the Euler equations:
\begin{equation}\label{eq2.12}
v_{\alpha\beta}^{\text{eff}}(r) = v_{\alpha\beta}(r)
+ \frac{\delta F^{ex}}{\delta\rho_\alpha({\bf r})} - \mu^{ex}_\alpha,
\end{equation}
where $v_{\alpha\beta}(r)$ is the direct interaction between species
and $\mu^{ex}_\alpha$ is the excess chemical potential. Thus the
electron-ion radial distribution function follows from the indirect
Kohn-Sham solution of the Euler-equation combined with the Percus
identity:
\begin{equation}\label{eq2.13a}
g_{eI}(r) = \frac{\rho_e^0(r\bigl|v^{\text{eff}}_{eI})}{\rho_e},
\end{equation}
where $\rho_e^0(r|v^{\text{eff}}_{eI})$ is the density of the unbound
(or free) valence electrons obtained from the wave functions that 
emerge from  a Kohn-Sham
solution of the one-center radially symmetric non-interacting
Schr\"{o}dinger equation in the external effective potential given by
(\ref{eq2.12})\cite{Kohn65}. The ion-ion radial distribution function
follows from a direct solution of the ion-ion Euler equation combined
with the Percus-identity:
\begin{equation}\label{eq2.13b}
g_{II}(r) =\frac{\rho_I(r\bigl|I)}{\rho_I} = \frac{\rho_I^0(r\bigl|v_{II}^{\text{eff}})}{\rho_I} = exp\left[-\beta v_{II}^{\text{eff}}(r)\right],
\end{equation}
where in the classical context $v_{II}^{eff}$ is commonly referred to
as the  {\em potential of mean force}.
Next we expand the effective potentials $v_{II}^{\text{eff}}(r)$ and
 $v_{eI}^{\text{eff}}(r)$ in a functional Taylor expansion around the
 equilibrium homogeneous densities and rewrite Eq. (\ref{eq2.12}) as:
\begin{equation}\label{eq2.13}
v_{\alpha I}^{\text{eff}}(r) = v_{\alpha I}(r)
   -\frac{1}{\beta}\sum_\gamma\rho_\gamma\int 
C_{\alpha\gamma}(|{\bf r}-{\bf r}^\prime|)
             h_{\gamma I}(r)d{\bf r}^\prime
+\frac{1}{\beta}B_{\alpha I}(r),
\end{equation}
where the $C_{\alpha \gamma}(r)$ are the homogeneous limits of Eq.\
 (\ref{eq2.6}), and the Percus trick was used to rewrite
 $(\rho_\gamma({\bf r}|v_{\gamma I}) - \rho_\gamma)$ in terms of the
 correlation functions $h_{\gamma I}(r)$ = $g_{\gamma I}(r) -1$.  The
 remaining third and higher order functional derivative terms are
 lumped into the so-called {\em bridge functions} $B_{\alpha
 \beta}(r)$. We note that these equations do not  hold for the effective
 electron-electron potential $v_{ee}(r)$.

Up to this point, our formulation is in principle exact.  However,
since the exact free-energy functionals and the related effective
external potentials are unknown, some approximations must be made.  In
the language of the theory of classical liquids\cite{Hans86}, we need
a {\em closure relation}.  For this we follow the approach developed
by Chihara, which he named the Quantum Hypernetted Chain Approximation
(QHNC)\cite{Chih78}.  The main approximations made by Chihara are
(roughly in ascending order of importance):
\begin{enumerate}

\item {\em The bare ion-ion potential is taken to be purely
Coulombic.}  This neglects core polarisation effects, but these are
expected to be small in the metals we study.  \item {\em The ion-ion
bridge function $B_{II}(r)$ is approximated by the one-component
bridge function of an appropriate reference state.}  This is commonly
called the RHNC or MHNC approximation\cite{Rose79}, and is expected to
be quite accurate. We use the repulsive part of the one-component
effective pair potential solved in the Percus-Yevick approximation as
a reference system to calculate the bridge function (see the Appendix
for details).

\item  {\em The electron-ion bridge function $B_{eI}(r)$ is set to $0$.} This
is commonly called the hypernetted chain (HNC) approximation, and is
generally also quite accurate, especially as the electron-ion correlations
are expected to be weaker than the ion-ion correlations.
\item {\em The local density approximation (LDA) is used for the
one-center electron-ion problem}.  The calculation of the electron-ion
correlation function reduces to calculating the Schr\"{o}dinger
equation in the external potential given by Eq.~(\ref{eq2.13}).  This
is similar to a self-consistent field all-electron calculation for a
single atom, except that the potential includes not only the nuclear
Coulomb contribution, but also terms reflecting the effect of the
surrounding ions.  We solve this effective atomic problem in the LDA,
which is widely used in electronic structure calculations.  The core
electrons are treated explicitly, i.e.\ this is an all-electron
calculation.  However, the core and valence screening effects are
separated in a manner similar to the linear unscreening procedure
used to derive pseudo-potentials\cite{Hafn87}.
\item{\em The valence electron correlations are treated in the jellium
approximation}. To calculate the full effective potentials, we need
the electron-electron direct correlation function $C_{ee}(k)$, which
can be re-written in terms of the so-called local field factors as was
done in Eq.~(\ref{eq2.10}) where the non-Coulombic correlation part
has been subsumed into the local field factor $G(k)$.  In the QHNC
approach, the local field factor is approximated to be that of jellium
at the average electron density, i.e.\ it is independent of ionic
correlations:
\begin{equation}\label{eq2.14}
C_{ee}(k) = -\beta v_{ee}(k)[1-G_{ee}^{jell}(k;\rho_e)].
\end{equation}
 Thus the electron-electron direct-correlation function uncouples from
the other correlation functions in Eq.~(\ref{eq2.8}).  This
approximation is similar in spirit to the LDA approximation and
greatly simplifies part of the electronic problem, but it is probably
the most serious and uncontrolled part of the QHNC closure.
\end{enumerate}

The approximations for the bridge-functions together with 
 Eqns. (\ref{eq2.13a}),
(\ref{eq2.13b}), (\ref{eq2.13}), and the closure
for $C_{ee}(k)$ in Eq.~(\ref{eq2.14}) 
 reduce the QOZ relations  of Eq.~(\ref{eq2.8})
 to a closed pair of  coupled
equations for the radial distribution functions:
\begin{equation}\label{eq2.15}
\rho_e g_{eI}(r) = \rho_e \Bigl(r\,\Bigl|\, v_{eI}(r)
-\frac{1}{\beta}\rho_I\int C_{e I}(|{\bf r}-{\bf
r}^\prime|) h_{I I}(r)d{\bf r}^\prime -
\frac{1}{\beta}\rho_e\int C_{e e}(|{\bf r}-{\bf r}^\prime|) h_{e
I}(r)d{\bf r}^\prime \,\,\Bigl)
\end{equation}
\begin{equation}\label{eq2.16}
g_{II}(r) = exp\Bigl(\,-\beta v_{II}(r) - \rho_I\int 
C_{I I}(|{\bf r}-{\bf r}^\prime|)
             h_{I I}(r)d{\bf r}^\prime - \rho_e\int 
C_{I e}(|{\bf r}-{\bf r}^\prime|)
             h_{e I}(r)d{\bf r}^\prime
+\frac{1}{\beta}B(r)\,\Bigl)
\end{equation}
which are solved self-consistently. This is the essence of the QHNC
approach: the original many-center problem has been reduced to an
effective one-center problem by replacing the many-body ion-ion
correlations with an  effective external potential that depends
self-consistently on the ion-ion correlations.  The main advantages
are that (a) no pseudo-potential is needed, i.e.\ it is an all-electron
calculation and (b) ion-ion and electron-ion correlations emerge
naturally and on the same footing.  Details of the (rather complex)
numerical implementation of the QHNC are described in the Appendix.

\section{Ion-Ion and Electron-Ion correlations}

\subsection{Ion-Ion radial distribution function}
\begin{figure}
\begin{center}
\epsfig{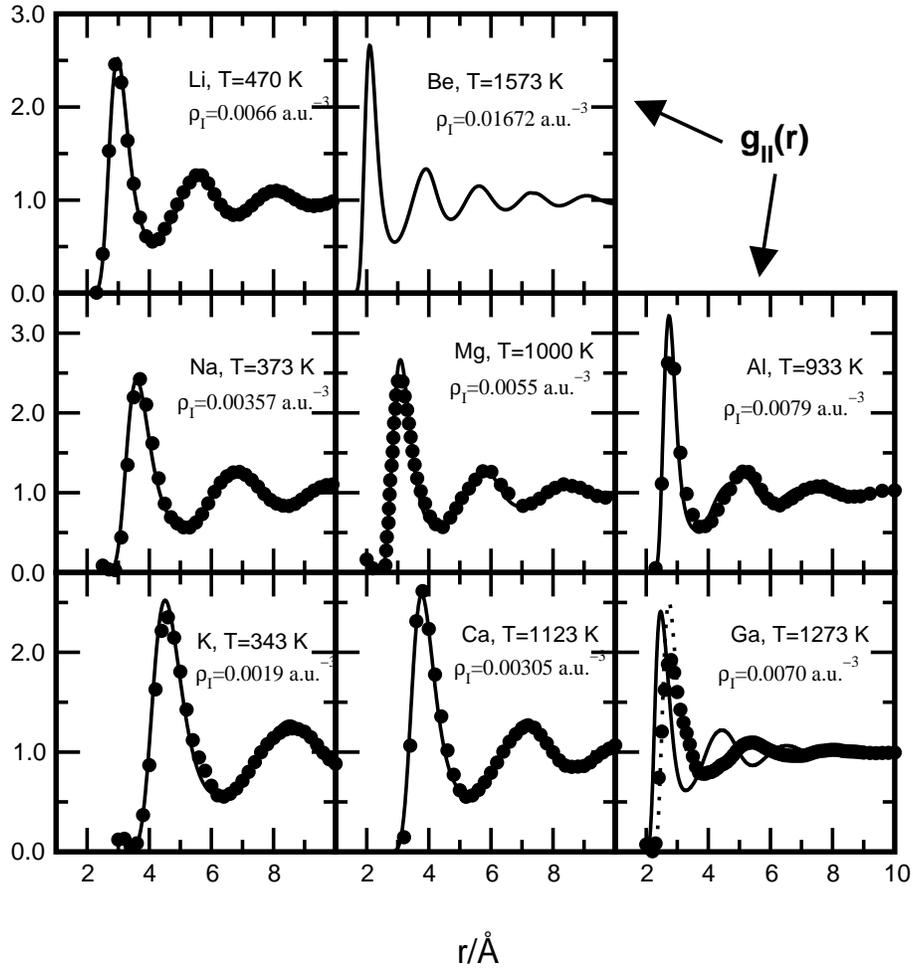} \vglue 0.1cm
\begin{minipage}{8cm}
\caption{Ion-ion radial distribution functions $g_{II}(r)$ calculated
by means of the QHNC method (solid lines) and compared to X-ray
experiments (circles)\protect\cite{Waseda}.  The dotted lines in the
Ga panel correspond to QHNC with the Ortiz-Ballone
$G(q)$\protect\cite{Orti94}. \protect\label{Fig1}}
\end{minipage}
\end{center}
\end{figure}

Armed with the QHNC approach, we can now tackle the electron-ion and
ion-ion correlation functions for a set of simple metals from the
first four rows of the periodic table. As a test of the approach, we
compare in Fig.~\ref{Fig1} the QHNC ion-ion radial distribution
function for our set of metals to the experimental X-ray data of the
Waseda group\cite{Waseda}. The QHNC provides a faithful representation
of $g_{II}(r)$ for all the metals except Ga.  The accuracy of the QHNC
for the other elements suggests that it can also be trusted for Be,
for which no experimental data could be found.

The case of Ga, however, calls for closer examination.  While the
exact form of $g_{II}(r)$ is sensitive to details of the liquid state
theory aspects of the closure, i.e.\ the form of the bridge function,
we tried various forms of the closure without much improvement.  On
the other hand, when we used the Ortiz-Ballone\cite{Orti94} local
field factor instead of the Ichimaru-Utsumi form\cite{Ichi81}, a
considerable improvement was obtained, in agreement with earlier
studies based on effective ion-ion potentials\cite{Boul97}. This
sensitivity of the QHNC approach to details of the local field factor
$G(q)$ suggests that approximation (5) of the previous section begins
to break down.  In addition, the d-electrons were very close to being
unbound, which made the QHNC algorithm difficult to converge. This
instability may be attributed to the implicit separation of the
exchange-correlation potential into bound and valence contributions in
approximation (4), i.e.\ the neglect of non-linear--core-corrections.
The fact that the Ortiz-Ballone $G(q)$ seems to work better for Ga is
most likely due to an accidental cancellation of errors.  It performs
considerably worse than the LDA or Ichimaru-Utsumi $G(q)$\cite{Ichi81}
for the other metals in our set.

\subsection{Pseudo-atoms and Electron-Ion correlations}

  The electron-ion structure factor, defined by Eq.~(\ref{eq1.1}),
can always be re-written in the following fashion:
\begin{equation}\label{eq3.5}
S_{eI}(k) = \frac{n(k)}{\sqrt{Z}} S_{II}(k),
\end{equation} which defines a new object, $n(k)$.  By taking the Fourier
transform we find, using Eq.~(\ref{eq1.4}), the electron-ion
radial-distribution function:
\begin{equation}\label{eq3.6}
\rho_e g_{eI}(r) = n(r) + \rho_i^0 \int_V n(r-r') g_{II}(r')
d{\bf r}',
\end{equation} which is proportional  to the probability of finding an
electron a distance $r$ away from an ion located at the origin.  Thus
a natural interpretation of $n(r)$ is the density of a
``pseudo-atom'', which, when superimposed according the ion-ion
radial-distribution function $g_{II}(r)$ gives the correct value of
the valence electron distribution.  The pseudo-atom is independent of
ionic correlations only to first order in the electron-ion potential,
at higher orders it implicitly includes 3-body and higher order ionic
averages\cite{Loui98b,Lai98,Loui98a}.

\begin{figure}
\begin{center}
\epsfig{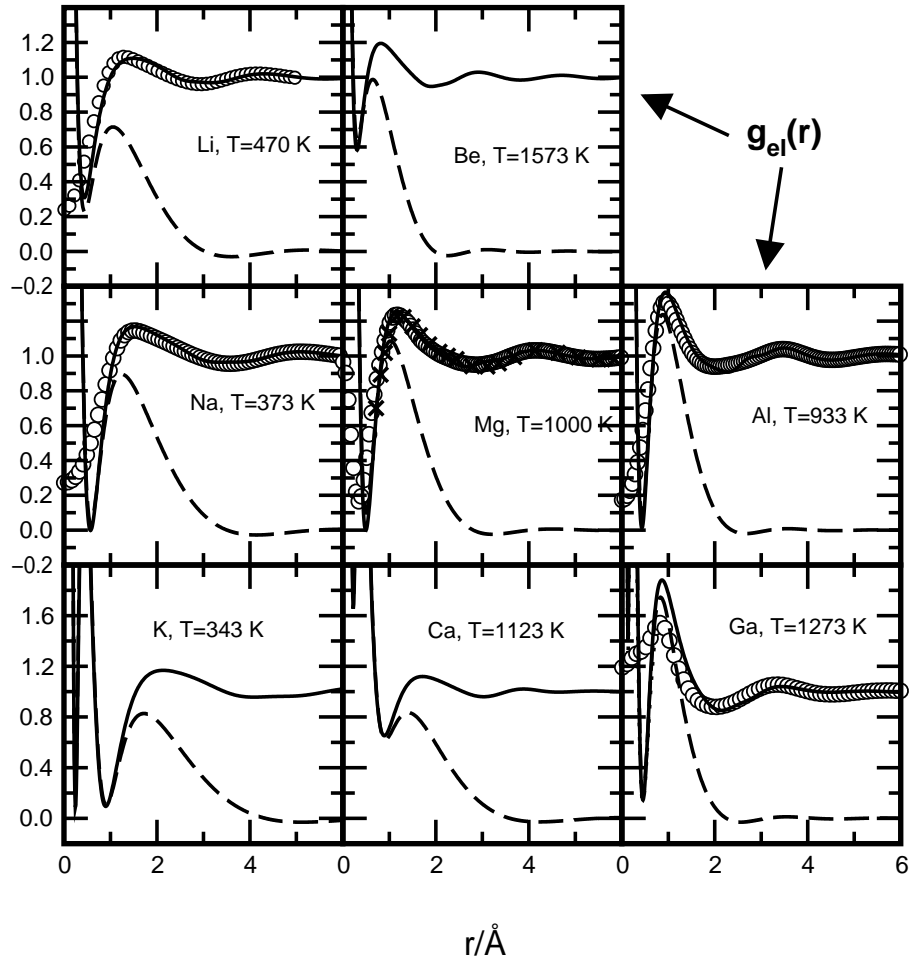} \vglue 0.1cm
\begin{minipage}{8cm}
\caption{Electron-ion radial  distribution functions as obtained from the QHNC
approximation (solid lines), the Orbital-free
method\protect\cite{Anta98,Anta99} (open circles) and Car-Parrinello
Molecular Dynamics\protect\cite{deWi95}(crosses). The dashed lines
represent the pseudo-atom density $n(r)/\rho_e$.
\protect\label{Fig2}}
\end{minipage}
\end{center}
\end{figure}

 In the QHNC approximation, the electron-ion radial-distribution
function follows directly from the solution of the one-body
Schr\"{o}dinger equation (Eq.~(\ref{eq2.13a})).  In Fig.~\ref{Fig2} we
show these electron-ion radial-distribution functions $g_{eI}(r)$ for
our set of simple metals.  Where possible, they have been compared to
{\em ab initio} Kohn-Sham\cite{Car85} and OF-AIMD\cite{Pear93}
results.  As is the case for the ion-ion radial distribution
functions, the QHNC approximation gives similar results to other
methods for all the elements except Ga, where once again an improved
agreement is obtained when the Ortiz-Ballone $G(q)$ is used.

It is instructive to compare the pseudo-atom density, included in
Fig.~\ref{Fig2} as $n(r)/\rho_e$, with the electron-ion radial
distribution function $g_{eI}(r)$. The pseudo-atom density goes to
zero for larger $r$, as it is essentially localised around a given
ion, while $g_{eI}(r)$ goes to 1 for large $r$, reflecting the fact
that outside the range of the ion's {\em own} pseudo-atom, $g_{eI}(r)$
simply probes the average density of the pseudo-atoms around the {\em
other} ions so that the probability of finding a valence electron a
distance $r$ away is simply related to the probability of finding an
ion there.  $g_{eI}(r)$ and $n(r)/\rho_e$ are essentially identical
for small $r$, as one might expect, while at larger $r$ the effect of
the ion-ion weighted superposition of the surrounding pseudo-atoms on
$g_{eI}(r)$ is evident.

Because $g_{eI}(r)$ implicitly includes a spherical average, all
angular bonding effects are effectively washed out, although an
indication of the effect of bonding can still be found by comparing
$g_{eI}(r)$ and a superposition of the free-atom electron
densities\cite{deWi95}.

\begin{figure}
\begin{center}
\epsfig{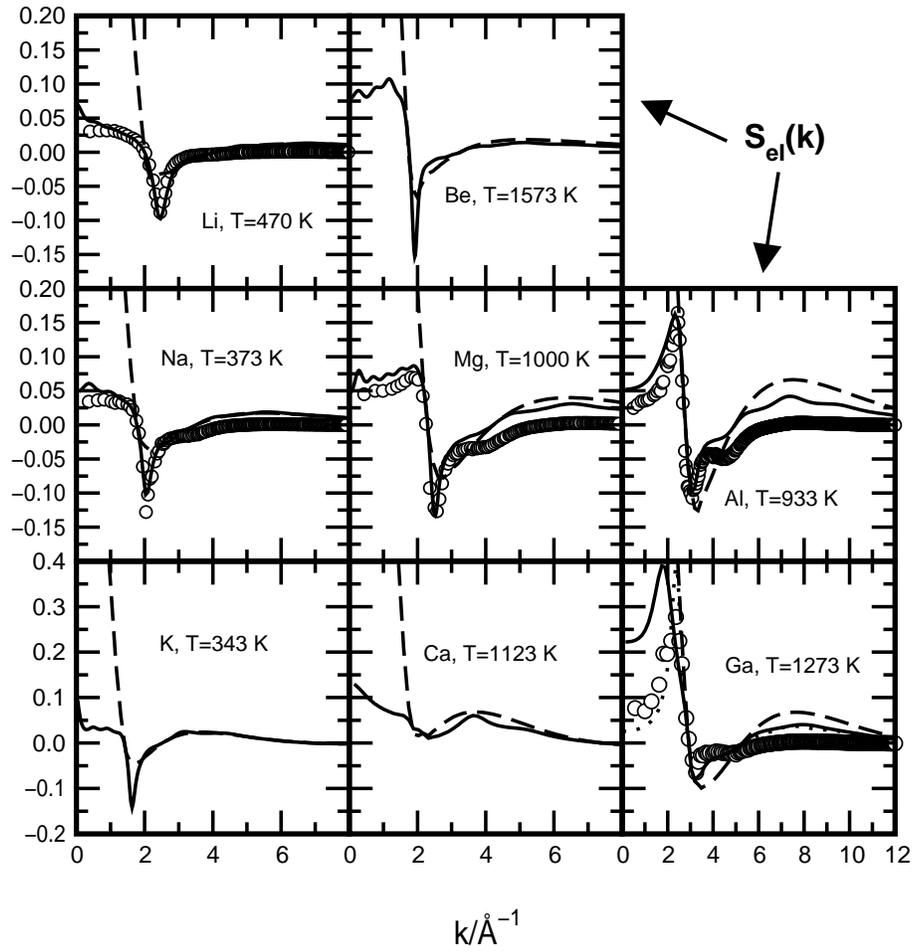} \vglue 0.1cm
\begin{minipage}{8cm}
\caption{Electron-ion structure factors $S_{eI}(k)$. The symbols have
the same meaning as in Fig.~\ref{Fig2}, the dashed lines again
represent the pseudo-atom density $n(k)$, but now in k-space. For the
scale, note that $n(k=0)=Z$, the number of valence electrons.
\protect\label{Fig3}}
\end{minipage}
\end{center}
\end{figure}

The relationships between the pseudo-atom, the ionic correlations, and
the electron-ion correlations become clearer in $k$-space where the
electron-ion structure factor is simply the product of the pseudo-atom
density and the ion-ion structure factor, as shown in Eq.\
(\ref{eq3.5}).  The ion-ion structure factor is sharply peaked at its
first maximum $k_p$ while the pseudo-atom density goes through zero at
$\bar{k}_0$.  If $\bar{k}_0 < k_p$, the product form implies that the
first peak of $S_{eI}(k)$ is negative, and the electron-ion structure
is in the so-called {\em low valence class}, while if $\bar{k}_0 >
k_p$, then the first peak of $S_{eI}(k)$ is positive, and the
electron-ion structure is in the so-called {\em high valence
class}\cite{Loui98b,Loui98a}.  In Fig.~\ref{Fig3} we plot both the
electron-ion structure factors $S_{eI}(k)$ and the pseudo-atom
densities $n(k)$ for our set of metals.  Li,
Be, Na, Mg and K are in the {\em low-valence class}, while Al and Ga
straddle the two classes.  Only Ca seems to fall outside this
taxonomy.

\section{Using Free-atom Form Factors  v.s. Metallic-Atom  Form Factors}

Neutron scattering probes the fluctuations of the nuclei, while X-ray
scattering probes the fluctuations of all the electrons.  In 1974,
Egelstaff {\em et.\ al.} \cite{Egel74} first suggested exploiting this
difference to extract electron-ion correlations for liquid metals.  In
1987, Chihara\cite{Chih87} re-examined the X-ray scattering problem,
demonstrating that a careful analysis of elastic and inelastic
contributions leads to the following coherent scattering
intensity\cite{Anta98}:
\begin{equation}\label{eq4.1}
I_X(k) = \bigl| f_{I}(k) + n(k) \bigl|^2 S_{N}(k),
\end{equation}
where $S_N(k)$ is the nucleus-nucleus structure factor which emerges,
for example, from neutron scattering, $f_I(k)$ is the ionic form
factor, i.e.\ the ionic electron density, and $n(k)$ is the
pseudo-atom density.  We shall call the object $f_M(k) = f_I(k) +
n(k)$ the {\em metallic-atom form factor}.  Traditionally the
structure factor from X-ray scattering  $S_X(k)$ has been extracted
from scattering intensity as follows:
\begin{equation}\label{eq4.2}
I_X(k) = \bigl| f_A(k) \bigl|^2 S_X(k),
\end{equation}
where $f_A(k)$ is the {\em free-atom form factor}, or the free-atom
electron-density.

\begin{figure}
\begin{center}
\epsfig{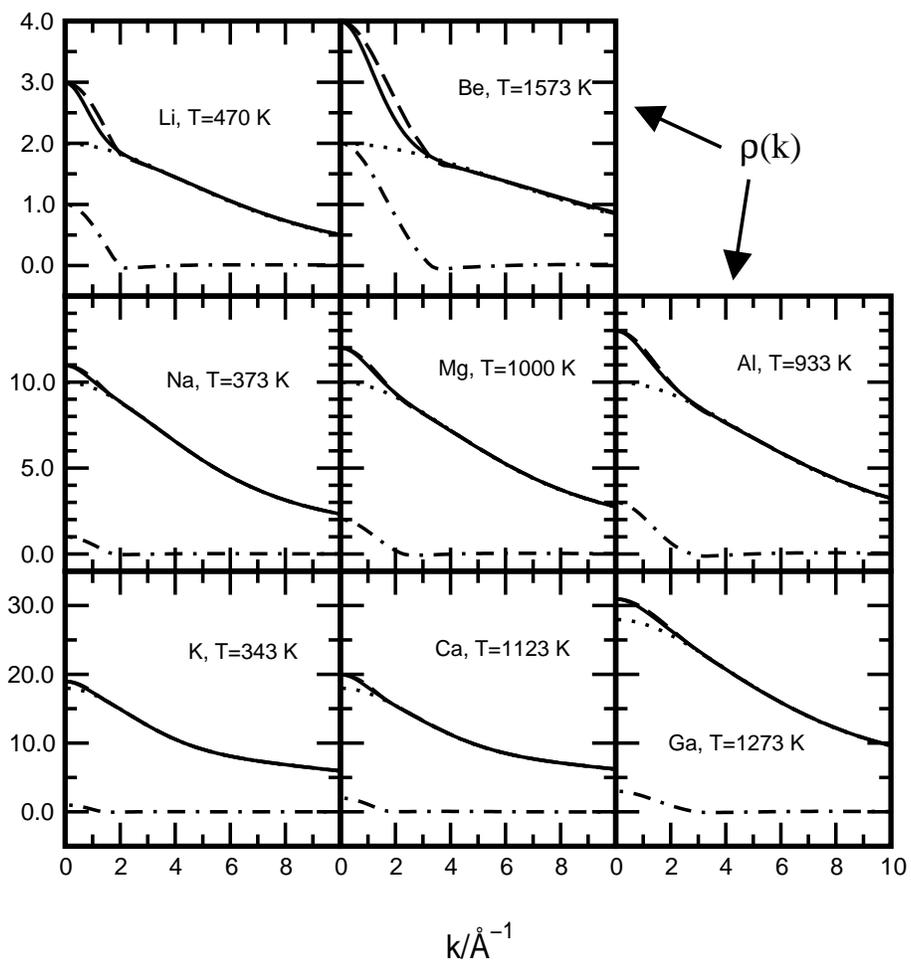} \vglue 0.1cm
\begin{minipage}{8cm}
\caption{Free-atom form factors $f_A(k)$ (solid lines), metallic-atom
form factors $f_M(k) = f_I(k) + n(k)$ (dashed lines), and ionic form
factors $f_I(k)$ (dotted lines), as predicted by the QHNC theory. The
chain lines represent the pseudo-atom density $n(k)$.
\protect\label{Fig4}}
\end{minipage}
\end{center}
\end{figure}

The difference between the two structure-factors, $S_N(k)$ and
$S_X(k)$, stems from the difference between the two form factors,
$f_A(k)$ and $f_M(k) = f_I(k) + n(k)$, and provides a measure of the
change in electron density upon bonding.  In Fig.~\ref{Fig4} we plot
the full free-atom (solid lines), metallic (dashed lines) and ionic
(dotted lines) form factors for our set of metals.  Also included are
the pseudo-atom densities (chain lines).  The ionic form factor is
essentially the same in the metallic and the free-atom environments,
so the difference between the metallic and free-atom form factors
stems from the difference between the pseudo-atom density and the
valence-electron density of the free atom.

Because  X-rays scatter off all the electrons, not just the valence
electrons, the effects of bonding are most pronounced when the ratio of the
number of valence electrons $Z$ to the total number of electrons
$Z_A$ is high.  Thus, as can be seen in Fig.~\ref{Fig4}, the effects
are largest in Li and Be, where the ratios $(Z:Z_A)$ are $(1:3)$ and
$(1:2)$ respectively, and the effect becomes smaller for the other
elements, where the ratios are Na: $(1:11)$, Mg: $(1:6)$, Al:
$(1:4.\bar{3})$, K: $(1:19)$, Ca: $(1:10)$, and Ga: $(1:10.\bar{3})$.

\begin{figure}
\begin{center}
\epsfig{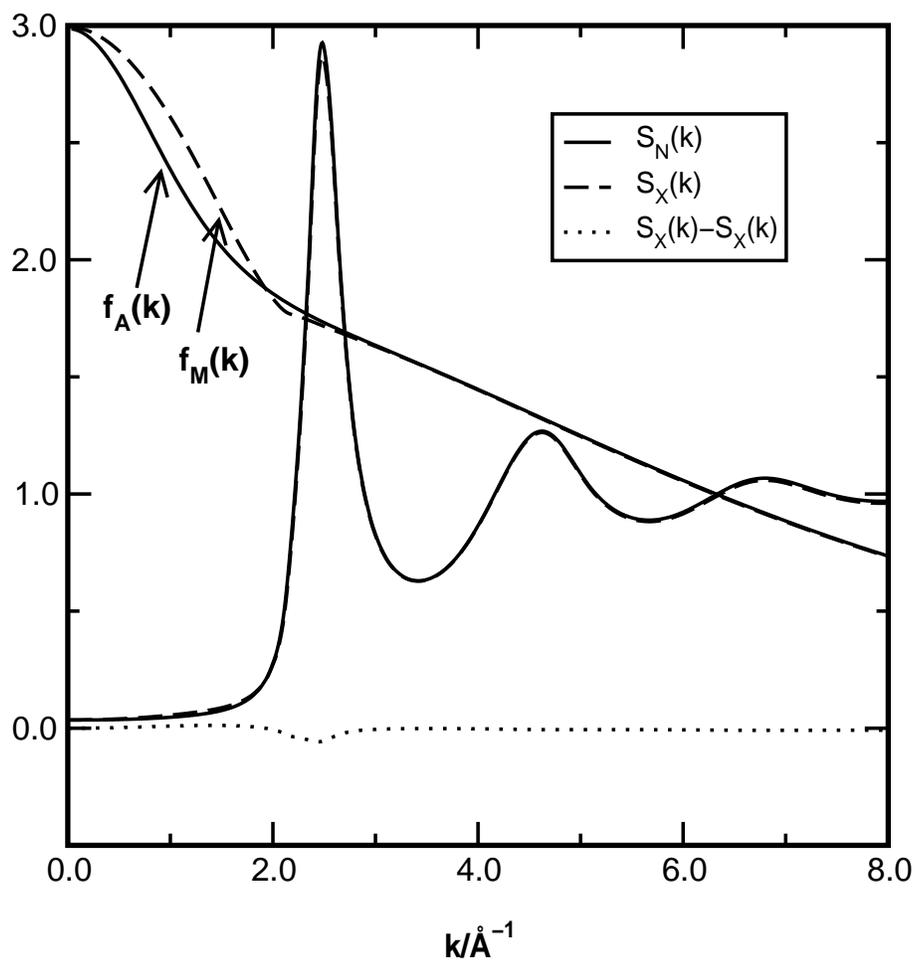} \vglue 0.1cm
\begin{minipage}{8cm}
\caption{The structure factors $S_N(k)$ (solid line) and $S_X(k)$
(dashed line), of liquid Li.  The dotted line corresponds to the
difference $S_X(k)-S_N(k)$. The metallic-atom and the free-atom form factor of
Li are also included in the figure.  \protect\label{Fig5}}
\end{minipage}
\end{center}
\end{figure}

\begin{figure}
\begin{center}
\epsfig{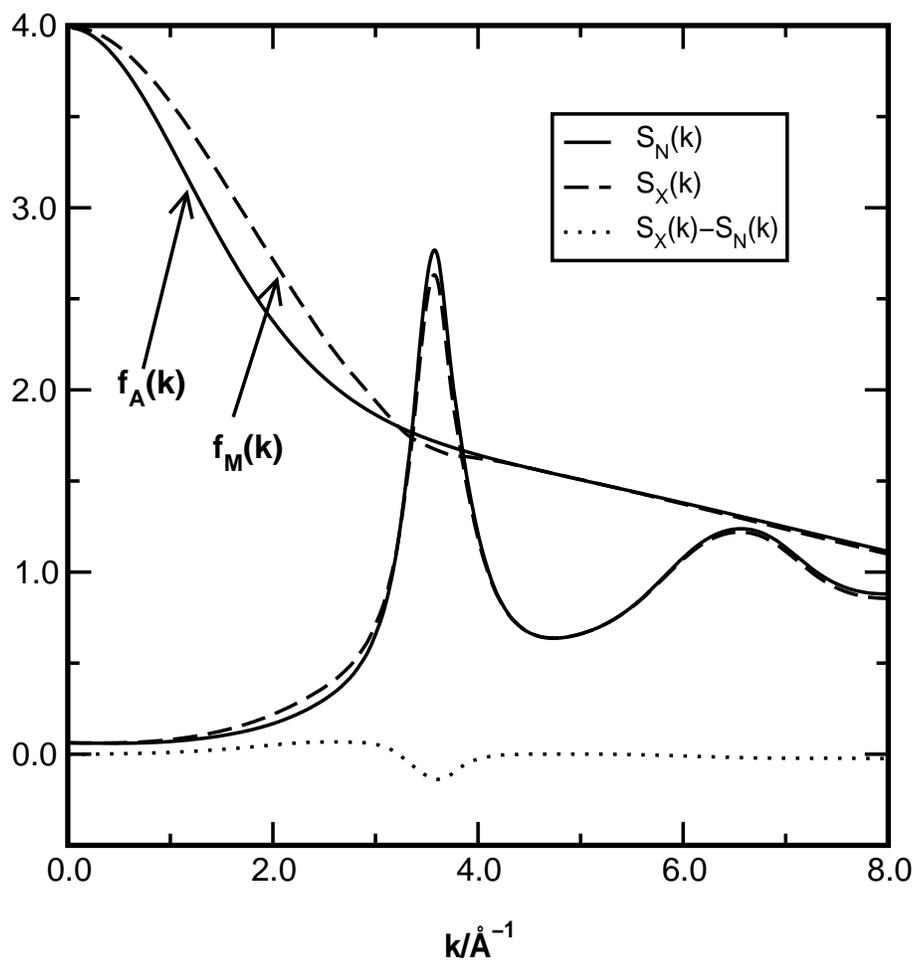} \vglue 0.1cm
\begin{minipage}{8cm}
\caption{Same symbols as in  Fig.~\protect\ref{Fig5}, but now for liquid Be.
\protect\label{Fig6}}
\end{minipage}
\end{center}
\end{figure}

In crystalline systems, X-ray studies of charge-densities only provide
information on the bonding density for certain fixed scattering
peaks. Similarly, in liquids  the scattering is strongest at the first
peak of the structure factor (at wave-number $k_p$), so to observe a
difference between $S_X(k)$ and $S_N(k)$ it is important that the
free-atom and the metallic-atom form factors differ near $k_p$.  This
is demonstrated for Li in Fig.~\ref{Fig5} and for Be in Fig.\
\ref{Fig6}.  Even though the difference between the free-atom and
metallic-atom  form factors is largest at $k$'s less than $k_p$, the
experimentally accessible difference, $S_X(k) - S_N(k)$, is largest at
$k_p$.

\begin{figure}
\begin{center}
\epsfig{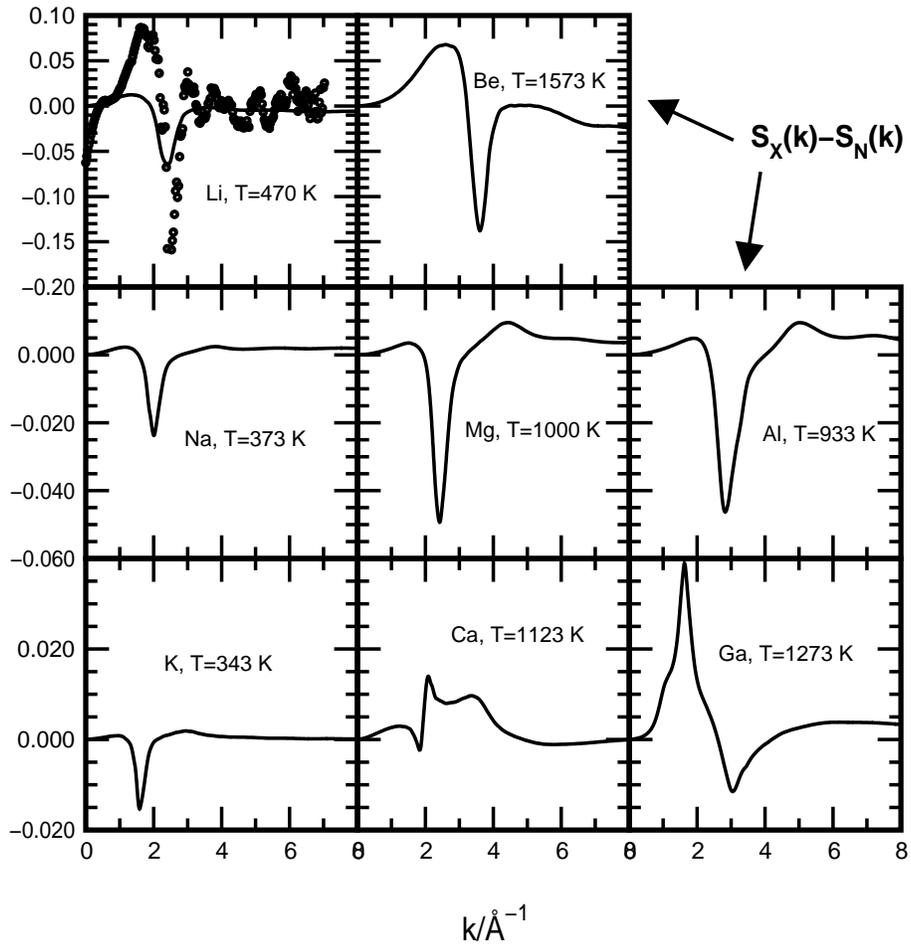} \vglue 0.1cm
\begin{minipage}{8cm}
\caption{Differences between using a free-atom and a metallic-atom
form factor to interpret X-ray scattering determinations of the static
structure factor $S_{II}(k)$  for a number of systems as
predicted by the QHNC theory (note the change in scale from 
panel to panel).  Alternatively, this can be viewed as
the difference between X-ray and neutron diffraction determinations of
the ion-ion structure factor.  Also included are the experimental
differences between X-ray and neutron diffraction for the ion-ion
structure factors of  Li from
reference [\protect\onlinecite{Olbr83}] \protect\label{Fig7}}
\end{minipage}
\end{center}
\end{figure}

In Fig.~\ref{Fig7} the difference, $S_X(k) - S_N(k)$, is shown for our
whole set of metals.  Generally the peak-height for $S_X(k)$ is
slightly lower than the peak-height for $S_N(k)$ and the two structure
factors are virtually identical away from $k_p$.  As was anticipated
in Ref.~\onlinecite{Loui98b}, the largest difference is for Be, where
$S_X(k_p)$ is about $5 \%$ lower than $S_N(k_p)$.  However, Be is
extremely toxic, and for that reason its static structure has not yet
been measured.  Perhaps the best chance of observing a difference
between $S_X(k)$ and $S_N(k)$ is for Li, Mg, or Al, where the
difference at $k_p$ is about $2 \%$.  Another possibility includes
liquid metallic Si, where the ratio is (1:3.5), and $\bar{k}_0$ is
expected to be greater than $k_p$ (i.e. Si's electron-ion structure
is expected to be in the high valence class), so that $S_X(k_p)$
is expected to be {\em larger} than $S_N(k_p)$ and the structure
factor may peak in a region where the two form factors differ by a
larger  amount than is the case for the low-valence class metals.

Measuring these differences will be extremely challenging, since they
require two completely different scattering techniques, which implies
subtracting two different sets of systematic corrections. In
particular, the removal of incoherent scattering effects from the
total scattering remains under discussion\cite{Olbr83,Chih87,Sinn97}.
We note that a series of experiments measuring the differences between
X-ray and neutron-scattering determinations of $S_{II}(k)$ have been
reported for Li\cite{Olbr83}, Na, Mg, Al, Zn, Ga, Sn, Te, Tl, Pb, and
Bi\cite{Take85}.  These measurements typically show differences that
are at best 5 to 10 times larger than expected from theoretical
treatments of the bonding effects, such as those shown for the QHNC in
Fig.~\ref{Fig7}\cite{Anta98}.  In fact for some of the heavier
elements, where the $S_X(k)-S_N(k)$ is expected to be very small due
to the large number of core electrons, the differences are several
orders of magnitude larger.  In Fig.~\ref{Fig7}, we include explicitly
the combined X-ray and neutron data of Olbrich {\em et.\
al.}\cite{Olbr83} for Li.  Even though their differences are smaller
than any of the differences measured in the other references cited in
[\onlinecite{Take85}] (in fact they are the only measurements which
fit within the scale of our graphs\cite{Anta98}), Olbrich {\em et.\
al.}  claim that experimental errors are too large to see bonding
effects in $S_{II}(k)$. For these reasons, the {\em interpretation} of
these measurements has been called into question by a number of
authors\cite{Chih89,Gonz93,Boul97,Anta98,Loui98b,Anta99}.  The
theoretical results are very robust, with simple linear response
theories in some cases agreeing quantitatively with the much more
sophisticated {\em ab-initio} Kohn-Sham calculations\cite{Loui98b}.
In a crystalline environment, the Kohn-Sham approach has been shown to
agree quantitatively to several significant figures with highly
accurate experimental measurements of the bonding
densities\cite{Lu93}, suggesting that the electron densities
calculated within the Kohn-Sham approach for the liquid state analogon
of these solid state measurements should be highly accurate as well.
In fact, for the Kohn-Sham type simulations, finite size and
statistical finite simulation time effects on the ion-ion structure
probably cause larger errors than errors arising from the
determination of the electron densities.  However, these simulation
errors are well understood, and will at most contribute a few {\em
relative} $\%$ to the difference $S_X(k) - S_N(k)$.  The
considerations above, coupled with the difficulties in dealing with
the subtraction of two very different sets of systematic corrections to
the data\cite{Salm99}, lead us to conclude that the experiments cited
have not yet attained an accuracy sufficient to measure the effects of
bonding in liquid metals.

 However, the advent of new high-accuracy X-ray and neutron beam
sources coming on line, together with the improvement of other
techniques such as anomalous X-ray scattering\cite{Pric98}, may bring
the measurement of these differences within experimental reach, at
least for a few of the metals in our set.  It seems increasingly
unlikely that this could be measured for many other elements where the
ratio $Z/Z_A$ is smaller and the core-electrons wash out any bonding
effects.

\section{Concluding Remarks}

We have carried out QHNC calculations for Li, Be, Na, Mg, Al, K, Ca,
and Ga.  The QHNC formalism, first introduced and mainly developed by
J. Chihara\cite{Chih76,Chih78,Chih85,Chih89} is a closure to the QOZ
relations.  Ion-ion and electron-ion correlations naturally emerge in
a unified fashion, and the interpretation of liquid metals in terms of
a ``pseudo-atom'' helps clarify the meaning of the electron-ion radial
distribution functions and structure factors.

The most serious approximation in the QHNC is probably approximation
(5) from section II.B, where the electron-electron direct-correlation
function $C_{ee}(k)$ is approximated by the form for jellium, making
it independent of the ion-ion and electron-ion correlations.  The
sensitivity to the local field factor $G_{ee}(k)$ found for Ga may
stem from a breakdown of approximation (5), but also from the neglect
of non-linear--core-corrections implicit in approximation (4).  Future
work will address both these issues.

The QHNC reduces to a linear-response formalism if the
direct-correlation function $C_{eI}(r)/\beta$ is approximated by its
low-density or long-range form $-v_{eI}(r)$, suggesting that the
accuracy of the QHNC probably benefits from an interference effect
which reduces the non-linear response terms\cite{Loui98a,Loui98b}.
For metallic hydrogen, where the lack of core-electrons implies no
interference effect, $C_{eI}(r)/\beta$ will differ significantly from
its low-density limit.  The relative importance of non-linear response
terms also suggests that approximation (5) may be poor for H.  In
addition, Xu {\em et.\ al.}\cite{Xu94} showed that small changes in
$C_{eI}(r)/\beta$ can have a large effect when input into DFT theories
of the freezing of monatomic H.  We expect the DFT theories to be
relatively less sensitive to changes in $C_{eI}(r)/\beta$ when applied
to the simple metals in our set.

The differences between X-ray measurements of the ion-ion structure
factor $S_{II}(k)$ interpreted with a free-atom or with a
metallic-atom form factor are the main experimentally relevant
quantities we calculate.  This difference, which reflects the effects
of metallic bonding of the valence electrons, is largest for elements
with a large ratio of valence to core electrons, such as Li, Be, Mg,
Al and maybe Si.  To date these bonding effects have not been
convincingly observed, but with new higher precision instruments
coming on line, they may soon be experimentally accessible.

\section{Acknowledgements}

We thank P. A. Madden, L.E. Gonz\'alez, P. Salmon, D. L. Price and
 M. L. Saboungi for helpful discussions, D. Rowan for a critical
 reading of the manuscript , and J. Chihara for help with some details
 of the implementation.  AAL thanks N. W.  Ashcroft for his insight in
 early stages of this work, and P. A. Madden for hospitality at
 Oxford, where some of this work was completed.  AAL also thanks the
 EC for support through the fellowship grant EBRFMBICT972464, and
 Hughes Hall, Cambridge, for a research fellowship. JAA thanks
 Ministerio de Educaci\'on y Cultura of Spain for the award of a
 postdoctoral fellowship in Oxford.

\section{Appendix:  Practical implementation of the QHNC approximation}

\subsection{Overview of the implementation}

In the practical implementation, we follow 2 steps to
self-consistency.

 {\bf Step 1: the ion-ion loop.} For a given $g_{eI}(r)$ and
$C_{eI}(r)$, an effective one-component ion-ion effective potential is
calculated and  the 1-component RHNC integral
equation is solved self-consistency for $g_{II}(r)$.

{\bf Step 2: the electron-ion loop.} For a given $g_{II}(r)$ and the
old $g_{eI}(r)$ and $C_{eI}(r)$, an effective electron-ion potential
$v_{eI}^{\text{eff}}(r)$ is calculated from Eq.~(\ref{eq2.13}).  The
self-consistent Schr\"{o}dinger equation is then solved to give a new
$g_{eI}(r)$ via Eq.~(\ref{eq2.13a}), and the procedure is repeated
to obtain self-consistency in $g_{eI}(r)$.

These two steps are then repeated until full self-consistency is obtained
between the two loops.

\subsection{Details of the the ion-ion loop}

We first rewrite the ion-ion problem as an effective one-component system
with the same radial distribution function:
\begin{equation}\label{eqA1}
g_{II}(r) = \exp[v_{II}^{\text{eff}}(r)] =  g(r) = \exp[v_1^{\text{eff}}(r)]
\end{equation}
where $v_{II}^{\text{eff}}(r)$ is the effective potential of mean force for
the ions, given by Eq.~(\ref{eq2.13}), and $v_1^{\text{eff}}(r)$ is the
effective potential of mean force for the one-component system.
The equality of the two radial-distribution functions then implies
that:
\begin{equation}\label{eqA2}
- \beta v_{II}(r) + h_{II}(r) - C_{II}(r) + B_{II}(r) = -\beta
  v_{1}(r) + h(r) - C(r)  + B(r),
\end{equation}
where $v_1(r)$ is the bare potential of the effective one-component
system, $C(r)$ is its direct correlation function, and $B_{II}(r)$
and $B(r)$ are the bridge functions of the two-component and
effective one-component systems respectively.  We follow Chihara and 
make the approximation\cite{bridge}:
\begin{equation}\label{eqA3}
B_{II}(r) \approx B(r),
\end{equation}
which, together with the QOZ relations of Eq.~(\ref{eq2.8}), implies that:
\begin{equation}\label{eqA4}
v_1(r) = v_{II}(r)    -\frac{\chi_{ee}^{(0)}(k)|C_{eI}(k)/\beta|^2}
                          {1+\chi_{ee}^{(0)}(k)C_{ee}(k)/\beta}.
\end{equation}
Note that if the electron-ion direct correlation function is replaced
by its low density or long-range limit: $C_{eI}(r)/\beta = -v_{eI}(r)$,
and Eq.~(\ref{eq2.14}) is used for $C_{ee}(k)$, the effective
one-component potential reduces to the usual linear screening
form\cite{Ashc78}.

Having now reduced the problem to an effective one-component form (by
assuming a fixed $g_{eI}(r)$ and $C_{eI}(r)$), we solve the
self-consistent RHNC equations in the usual
way\cite{Hans86,Lomba89,Martin93}.  The bridge function is obtained
using as a reference system the repulsive part of the $v_1(r)$ solved in
the Percus-Yevick approximation\cite{rRHNC} which is known to perform
well for short-range potentials\cite{Hans86}. The bridge function
obtained in this way gives very similar results to the standard RHNC
approximation (where the reference system is the hard-sphere fluid)
for the systems here studied but with the advantage that no
optimization of the hard-sphere diameter is required. This feature is
specially recommended in the context of the QHNC theory, where the
RHNC equation is solved in combination with the ion-electron integral
equation.

\subsection{Details of the electron-ion loop}

For a given $g_{II}(r)=g(r)$, and an old $g_{eI}(r)$ and  $C_{eI}(r)$,
the new effective electron-ion potential follows from
Eq.~(\ref{eq2.13}):

\begin{equation}\label{eqA5} v_{eI}^{\text{eff}}(r) = v_{eI}(r) -
\rho_e\int \frac{C_{e e}(|{\bf r}-{\bf r}^\prime|)}{\beta} h_{e
I}(r)d{\bf r}^\prime \,
-\rho_I\int \frac{C_{e I}(|{\bf r}-{\bf
r}^\prime|)}{\beta} h_{I I}(r)d{\bf r}^\prime, 
\end{equation} where the first term, $v_{eI}(r)$, is the bare electron-ion
interaction, the second term describes the screening by the valence
electrons, both those of the central ion as well as those originating
from the pseudo-atom densities of the surrounding ions, and the third
term describes the interaction of the electrons with the other ions.
For the local-field factors implicit in $C_{ee}(k)$, we used the
Ichimaru-Utsumi form\cite{Ichi81}, but except for Ga, the simpler LDA
form also performed quite well.  For the bare electron-ion interaction
we follow Chihara\cite{Chih85} and write:
\begin{equation}\label{eqA6} v_{eI}(r) = -\frac{Z_Ae^2}{r}
+ \int v_{ee}(|{\bf r}-{\bf r}^\prime|)\rho^b_e(r)d{\bf r}^\prime +
\mu_{XC}[\rho^b_e(r)+\rho_e] - \mu_{XC}[\rho_e], 
\end{equation} where
$Z_A$ is the nuclear charge, $\mu_{XC}[\rho(r)]$ is the
exchange-correlation part of the free energy functional (we take the
usual LDA parameterisation of Perdew and Zunger\cite{Perd81} of the
Ceperley-Alder quantum monte-carlo simulations\cite{Cepe80}), and
$\rho^b({\bf r})$ is the bound electron density obtained from the
solution of the Schr\"{o}dinger equation.  This form is not exact
within the LDA, as its derivation implies a linear unscreening
process, neglecting the so-called
non-linear--core-corrections\cite{Hafn87}.  In fact, this linear
unscreening process is not necessary, and the full screening from the
combined valence and core electron densities can be taken into
account, but this will be addressed in a later publication.

Using this effective electron-ion potential, the 
one-electron Schr\"{o}dinger  equation:
\begin{equation}\label{eqA7}
\left[-\frac{\hbar^2}{2m_e}\nabla^2+v_{eI}^{\text{eff}}(r)\right]
\psi^i_e(r) = \epsilon^i_e\psi^i_e(r),
\end{equation}
is solved for the effective potential of Eq.~(\ref{eqA5}).  The bound
electron density is then calculated by means of
\begin{equation}
\label{sumdens}
\rho^b_e({\bf r}|v_{eI}^{\text{eff}}) = \rho^b_e({\bf r}) = 
\sum_{i(b)}\left|\psi^{i(b)}_e(r)\right|^2 
\end{equation}
where the index $(b)$ refers to {\em bound} states, while
$\rho_e^{f}(r)$, the {\em unbound} density directly related to
$g_{eI}(r)$ through Eq.(\ref{eq2.13a}) corresponds to the continuum
part of the eigenvalue spectrum (positive energies) and is calculated
as a superposition of {\em scattering states}. In atomic units this is
given by \cite{Dhar82,Dage72}:
\begin{equation}
\label{continuum}
\rho_e^{f}(r|v_{eI}^{\text{eff}}) = \rho_e + \frac{1}{\pi^2}\int_0^{k_F} 
dk\;k^2 \sum_l (2l+1)\left[R_{kl}^2(r)-j_l^2(rk)\right]
\end{equation}
where $k_F$ is the Fermi wave vector and $R_{kl}(r)$, the radial part of the 
wave function, is a solution of the equation:
\begin{equation}
\label{radial}
\frac{d^2(rR_{kl})}{dr^2} +
\left[k^2-\frac{l(l+1)}{r^2}-2v_{eI}^{\text{eff}}(r)\right] rR_{kl}(r)
= 0.
\end{equation}
$R_{kl}(r)$ must be normalized by its asymptotic limit, i.e.\
\begin{equation}
\label{asymtotic}
\lim_{r\rightarrow\infty} \left[rR_{kl}(r)\right] = 
	j_l^2(rk)\cos\eta_{l}(k) + n_l^2(rk)\sin\eta_{l}(k),
\end{equation}
where $j_l(x)$ and $n_l(x)$ are  spherical Bessel functions and 
$\eta_{l}(k)$ is the {\em phase shift}. The phase shifts at the Fermi level 
fulfil 
the {\em Friedel sum rule}\cite{Chih85}:
\begin{equation}
\label{Friedel}
\frac{2}{\pi}\sum_l(2l+1)\eta_{l}(k_F) = ZS_{II}(0)
\end{equation}
where $Z$ is the ionic charge and $S_{II}(0)$ the long-wavelength limit of
the ion-ion structure factor.

	We solve the Schr\"{o}dinger equation in two stages: 

{\bf (1)} We first look
for bound states and the eigenvalues of Eq.~(\ref{eqA7}) using the
predictor-corrector method on a logarithmic grid\cite{Troullier}.

{\bf (2)} Once the bound density is computed (needed to obtain the
electron-ion bare interaction via Eq.~(\ref{eqA6})), we solve
Eq.~(\ref{radial}) for scattering states on a Hermann-Skillmann mesh
using the Numerov method. Both the logarithmic and the
Hermann-Skillmann grids, as well as the linear mesh in which the
correlation functions are stored, span the same range of 40.96
a.u. (see below for numerical details). We follow Chihara and
introduce a cutoff radius $r_{\text{cut}}$ outside which the solutions
are taken to be of the usual Friedel oscillatory form:
\begin{equation}
\label{osci}
v_1(r) \propto \frac{1}{4\pi r^3}\cos(2k_Fr) $\hspace{1cm}$ r >
r_{\text{cut}},
\end{equation}
where $r_{\text{cut}}$ is typically equal to around $4-5$ times the
electron Wigner-Seitz radius. This procedure avoids difficulties with
the long-range nature of the potentials while keeping the number of
mesh-points needed at a manageable level. Typically, we needed 4 or 5
iterations of the ion-ion and electron-ion loops for a maximum
tolerance of 0.1\% error between successive solutions to converge.
The screening relation
\begin{equation}
\label{screening}
\int \rho_e^{f}(r)d{\bf r} = Z
\end{equation}
is fulfilled with an error around $0.1 \%$ for the lower valence
metals, and less than $1 \%$ for Ga or Al. Although this is somewhat
unsatisfactory, and is a larger error than that reported in previous
studies\cite{Dhar82,Dage72}, we found that it does not affect the
shape of the correlation functions or the effective pair potential at
short and intermediate distances.  Future work will address this issue
in more detail.

	We use a linear grid of 4096 points for the ion-ion and
electron-ion correlation functions (i.e.\ a grid size of 0.01 a.u. for
a maximum distance of 40.96 a.u.). The logarithmic grid contains 983
points whereas the Hermann-Skillmann grid comprises 5 blocks of
typically 80, 40, 40,  90, and 800 points respectively. This
corresponds to a grid size of 0.0015 for the first block. The
Schr\"{o}dinger equation for scattering states is solved for 25
equally-spaced $k$-points between 0 and $k_F$ and up to 11 plane waves
($l_{max}=10$). Then, the integral in Eq.~(\ref{continuum}) is
computed using Simpson's rule with the same number of $k$-points. An
interpolation by cubic splines is employed to swap between the
different grids involved in the calculation (linear for correlation
functions and logarithmic and Hermann-Skillmann grids to solve the
Schr\"{o}dinger equation). Fast Fourier transforms are used to convert
correlation functions from real to reciprocal space. In addition, we
have implemented Ng's method\cite{Ng74} in the ion-ion and the 
ion-electron loops  to accelerate the convergence.

\subsection{Details of the initial setup}

	As mentioned before, the implementation of this iterative
procedure requires an initial effective pair potential.  Following
Chihara\cite{Chih89}, we make use of the so-called {\em
jellium-vacancy model} (JVM) to obtain such initial potential. The JVM
can be derived directly from the QHNC approach by the following 2
approximations:
\begin{enumerate}
\item The ion-ion correlation function approximated as a  step
function:
\begin{equation}
\label{jvm1}
h_{II}(r) = \left\{
\begin{array}{rcc}
 -1 & \mbox{for} & r < R \\
  0 & \mbox{for} & r > R 
\end{array}
\right.
\end{equation}
where $R$ is the ion Wigner-Seitz radius, i.e.,
$R=[3/(4\pi\rho_I)]^{1/3}$.
\item The electron-ion DCF used in the effective electron-ion
interaction of Eq.~(\ref{eqA5}) is approximated to be of a purely Coulombic
form:
\begin{equation}
\label{jvm2}
C_{eI}(r)/\beta = \frac{Z_Ie^2}{r}.
\end{equation}
\end{enumerate}
These two approximations result in an 
electron-ion  potential $v_{eI}^{\text{eff}}(r)$ which is independent
on the ion-ion correlations\cite{Chih89}. 
The Schr\"{o}dinger equation that follows from this effective potential
can then be solved self-consistently as described above, and the
ensuing $g_{eI}(r)$ and new $C_{eI}(r)$ used to derive an effective
ion-ion pair-potential from Eq.~(\ref{eqA4}), for use in the 
initial ion-ion loop.

\newpage




\begin{thebibliography}{99}
\bibitem[*]{Juan} present address: Department of Chemistry, Imperial
College of Science, Technology and Medicine, Exhibition Road, London,
SW7 2AY, UK.
\bibitem{foot1}Compared to the classical liquid, electron liquids are
more weakly correlated (see e.g. N.W. Ashcroft in {\em Density
Functional Theory}, edited by E. K. U.~Gross and R. M.~Dreizler
(Plenum Press, New York 1995).), which is why we refer to them as
fluids instead of liquids.
\bibitem{Ashc78} N. W.~Ashcroft and D.~Stroud, Solid State Physics
{\bf 33}, 1 (1978); R.~Evans in {\em Electrons in Disordered Metals
and at Metal Surfaces}, edited by P. Phariseau, B.L. Gyorffy and L. Scheire,
Plenum, ASI Series B {\bf 42}, (1979).
\bibitem{Loui98b} A. A. Louis and N. W. Ashcroft,
Phys. Rev. Lett. {\bf 81}, 4456 (1998); see also, A. A. Louis and
N. W. Ashcroft, J. Non-Cryst. Solids, {\em to appear} (1999).
\bibitem{Cusa76} S. Cusack, N. H. March, M. Parrinello and M. P. Tosi,
J. Phys F: Met. Phys. {\bf 6}, 749 (1976).  K. Hoshino and M. Watabe,
J. Phys. Soc. Japan {\bf 61}, 1663 (1992).
\bibitem{Boul97} M. Boulahbak, J. F. Wax, N. Jakse, and
J. l.~Bretonnet, J. Phys.: Condens. Matter, {\bf 9}, 4017 (1997).
\bibitem{Wax97} J.~F.~Wax, N.~Jakse, and J.~L.~Bretonnet,
Phys. Rev. B {\bf 55}, 12099 (1997).
\bibitem{Marc98} N. H. March and M. Tosi, Laser and Particle Beams,
{\bf 16}, 71 (1998); N.~H.~March, Current Science, {\bf 75}, 1246
(1998).
\bibitem{Lai98}S.~K.~Lai, K.~Horii and M~Iwamatsu, Phys. Rev. E {\bf
58}, 2227 (1998).
\bibitem{Loui98a} A. A. Louis, PhD thesis, Cornell University (1998)
(unpublished -- available at
http://ket.ch.cam.ac.uk/people/ardlouis/Thesis/PhDthesis.html).
\bibitem{Anta98} J. A. Anta, B. J. Besson, and P. A. Madden,
 Phys. Rev. B {\bf 58}, 6124 (1998).
\bibitem{Kohn65} W.~Kohn and L. J.~Sham, Phys. Rev. {\bf 140}, A1133
(1965).  
\bibitem{Car85} R. Car and M. Parrinello,
Phys. Rev. Lett. {\bf 55}, 2471 (1985).
\bibitem{deWi95} G. A. de Wijs, B. Pastore, A. Selloni, and W. van der
Lugt, Phys. Rev.  Lett. {\bf 75}, 4480 (1995).
\bibitem{Pear93} M. Pearson, E. Smargiassi, and P. A. Madden,
J. Phys.: Condens. Matter, {\bf 5}, 3221 (1993).
\bibitem{Anta99} J. A.  Anta and P. A. Madden,
J. Phys.: Condens. Matter, {\em to appear} (1999).
\bibitem{Chih78} J. Chihara, Prog. Theor. Phys. {\bf 59}, 76 (1978).
\bibitem{Xu98} H. Xu and J.-P. Hansen, Phys. Rev. E {\bf 57}, 211
(1998).
\bibitem{Merm65} N. D. Mermin Phys. Rev. {\bf 137}, A1441 (1965).
\bibitem{Zinn89} J.~Zinn-Justin, {\em Quantum Field Theory and
Critical Phenomena}, (Oxford University Press, Oxford (1989)).
\bibitem{foot2} This is actually the definition of $C({\bf r}, {\bf
r'})/\beta$ (where $\beta^{-1} = k_B T$) , a notation that stems from
the classical context.  At zero-temperature, only this ratio is well
defined and given by the relation~(\ref{eq2.6}); $C({\bf r}, {\bf
r'})$ itself is not.
\bibitem{Orns14} L.S. Ornstein  and F. Zernike, {\em Proc. Akad. Sci.} 
(Amsterdam){\bf 17}, 793 (1914).
\bibitem{Hans86} J.-P.~Hansen and I. R.~McDonald, {\em Theory of
Simple Liquids, 2nd Ed.}, (Academic Press, London (1986)).
\bibitem{Kubo64} R. Kubo, Rep. Prog. Phys. {\bf 19}, 255 (1966).
\bibitem{Ashc76} N. W.~Ashcroft and N .D.~Mermin, {\em Solid State
Physics} (Holt, Rinehart and Winston, New York 1976).
\bibitem{Chih76} J. Chihara, Prog. Theor. Phys. {\bf 55}, 340 (1976).
\bibitem{Ichi85} S. Ichimaru, S. Mitake, S. Tanaka and X-Z. Yan, Phys.
Rev. A {\bf 32}, 1768 (1985).
\bibitem{Perc62} J. K.~Percus, Phys. Rev. Lett. {\bf 8}, 462 (1962);,
see also the appendix of J. Chihara, J. Phys.: Condens. Matter {\bf
3}, 8715 (1991).
\bibitem{Rose79} Y. Rosenfeld and N.W. Ashcroft, Phys. Rev. A {\bf
20}, 1208 (1979)
\bibitem{Hafn87} J.~Hafner, {\em From Hamiltonians to Phase Diagrams},
(Springer Verlag, Berlin, (1987)).
\bibitem{Waseda}   IAMP database of [SCM-LIQ], http://www.iamp.tohoku.ac.jp.
\bibitem{Orti94} G. Ortiz and P. Ballone, Phys. Rev. B. {\bf 50}, 1391
(1994).
\bibitem{Ichi81} S. Ichimaru and K. Utsumi, Phys. Rev. B. {\bf 24},
7385 (1981).
\bibitem{Egel74} P. A. Egelstaff, N. H. March and N. C. McGill,
Can. J.  Phys. {\bf 52}, 1651 (1974).
\bibitem{Chih87} J. Chihara, J. Phys. F: Met. Phys. {\bf 17}, 295 (1987).
\bibitem{Olbr83} H. Olbrich, H. Ruppersberg, and S. Steeb,
Z. Natur. A, {\bf 38}, 1328 (1983).
\bibitem{Sinn97} H. Sinn, {\em et.\ al.}, Phys. Rev. Lett. {\bf 78}, 1715
(1997).
\bibitem{Take85} S. Takeda, S. Tamaki and Y. Waseda, J. Phys. Soc.
Japan {\bf 54}, 2552 (1985)(Bi, Sn); S. Takeda, S. Harada, S. Tamaki
and Y.  Waseda, J. Phys. Soc.  Japan {\bf 55}, 184 (1986)(Zn, Pb);
S. Takeda {em et.\ al.}  {\em ibid} {\bf 55}, 3437 (1986)(Ga, Tl), {\em ibid}
{\bf 58}, 3999 (1989) (Na), {\em ibid} {\bf 60}, 2241 (1991) (Al),
{\em ibid} {\bf 62}, 4277 (1993) (Te), {\em ibid} {\bf 63}, 1794
(1994) (Mg), S. Takeda {\em et.\ al.}, J. Non-Cryst. Solids {\bf 207}, 365
(1996) (Na, Mg, Al).
\bibitem{Gonz93} L.E. Gonzalez, D.J. Gonzalez, and K. Hoshino,
J. Phys.: Condens. Matter {\bf 5}, 9261 (1993).
\bibitem{Lu93} Z.W. Lu, A. Zunger and M. Deutsch, Phys. Rev. B {\bf
47}, 9385 (1993).
\bibitem{Salm99} For example, small differences in standard recipes
for computing resolution functions in neutron diffraction easily cause
differences in the first $S_{II}(k)$ peak of order 1 to 2 \%, roughly
equal to the effect of bonding for Li, Mg or Al,  P. Salmon, {\em
private communication}.
\bibitem{Pric98} D.L. Price, M.L. Saboungi and A.C. Barnes,
Phys. Rev. Lett. {\bf 81}, 3207 (1998).
\bibitem{Chih85} J. Chihara,  {\em J. Phys. C.: Solid State Phys} {\bf
18}, 3103 (1985).
\bibitem{Chih89} J. Chihara, Phys. Rev. A {\bf 40}, 4507 (1989);
M. Ishitobi and J. Chihara, {\em J. Phys.: Condens. Matter} {\bf 4},
3679 (1992).
\bibitem{Xu94} H. Xu, J.P. Hansen, and D. Chandler,
Europhys. Lett. {\bf 36}, 419 (1994).
\bibitem{bridge} The implications of this approximation are discussed
in chapter 6 of Ref. [\onlinecite{Loui98a}].
\bibitem{Lomba89} E. Lomba, {\em Mol. Phys.} {\bf 68}, 87 (1989)
\bibitem{Martin93} C. Mart\'{\i}n, E. Lomba, J. A. Anta and
M. Lombardero, {\em J. Phys.: Condens. Matter} {\bf 5}, 379 (1993)
\bibitem{rRHNC} L. E. Gonz\'alez, {\em private communication}.
\bibitem{Perd81} J.P. Perdew and A. Zunger, Phys. Rev. B. {\bf 23},
5048 (1981).
\bibitem{Cepe80} D.M. Ceperley and B. Alder, Phys. Rev. Lett. {\bf 45},
566 (1980).
\bibitem{Dhar82} M. W. C. Dharma-Wardana and F. Perrot, {\em
Phys. Rev. A} {\bf 26}, 2096 (1982).
\bibitem{Dage72} L. Dagens {\em J. Phys. C: Solid State Phys} {\bf
5}, 2333 (1972).
\bibitem{Troullier} For this purpose we have modified a FORTRAN code
by N. J. Troullier and J. L. Martins that solves the Kohn-Sham
equation for atoms (University of Minnesota, Sept, 1989; original by
S. Froyen, Berkeley).
\bibitem{Ng74} K. Ng, {\em J. Chem. Phys.} {\bf 61}, 2680 (1974).



\end{thebibliography}
\end{document}